\documentclass[12pt, a4paper]{article}
\pdfoutput=1

\usepackage{amsmath}
\usepackage{amsfonts}
\usepackage{amssymb}
\usepackage{latexsym}
\usepackage{graphicx}
\usepackage{color}
\usepackage{mathrsfs}
\usepackage{slashed}
\usepackage{hyperref}
\usepackage{datetime}

\numberwithin{equation}{section}

\hypersetup{colorlinks, citecolor=bluscuro, linkcolor=bluscuro, urlcolor=bluscuro}
\definecolor{rossos}{rgb}{0.8,0.2,0.3}
\definecolor{bluscuro}{rgb}{0.15, 0.2, 0.9}
\definecolor{verdes}{rgb}{0.1, 0.5, 0.1}

\makeatother   

 \def\be   {\begin{equation}}   \def\ee   {\end{equation}}
 \def\ba   {\begin{array}}      \def\ea   {\end{array}}
 \def\bea  {\begin{eqnarray}}   \def\eea  {\end{eqnarray}}
 \def\bean {\begin{eqnarray*}}  \def\eean {\end{eqnarray*}}

\begin{document}


\vspace{7cm}

\begin{center}
\vspace{5cm}
{\Large \textsc {
Higgs decay into photons through a spin-2 loop
}}
\\ [1.5cm]
{ \textsc{
Alfredo Urbano
}}
\\[1cm]

\textit{
Laboratoire de Physique Th\'eorique de l'\'Ecole Normale Sup\'erieure\\ \vspace{1.5mm}
24 rue Lhomond, F-75231 Paris, FRANCE.
}
\\ [3.5cm]
{ \textsc{
Abstract
}}
\\ [1.5cm]
\end{center}

A new particle with proprieties  similar to those of the Higgs boson in the Standard Model (SM) has been recently discovered. The biggest discrepancy is related to its diphoton decay, whose branching ratio seems to be around two times larger with respect to the correspondent SM value; this evidence, even if still affected by large uncertainties, suggests that clues of new physics related to the spontaneous breaking of the electroweak symmetry could be hidden under this loop-induced process. A new strongly-coupled sector responsible for this breaking, for instance, could produce in analogy with QCD a charged massive spin-2 state. In light of these arguments we calculate and discuss the role of such a resonance in the diphoton decay width of the Higgs.


\def\thefootnote{\arabic{footnote}}
\setcounter{footnote}{0}
\pagestyle{empty}

\newpage
\pagestyle{plain}
\setcounter{page}{1}

\section{Introduction}

Recently both ATLAS \cite{Giannotti,:2012gk} and CMS \cite{Incandela,:2012gu} experiments, presenting at CERN their analysis based on the data collected in 2012 by the Large Hadron Collider (LHC), announced the discovery of a new  particle  with proprieties remarkably similar to those of the Higgs boson \cite{Higgs}  in the Standard Model (SM). This astonishing discovery represents the greatest achievement in the history of  high-energy physics, marking out a bright milestone in our understanding of the Universe:  from the long-standing Higgs hunting era we are now entering into the era of  Higgs precision measurements.\\
To truly understand the mechanism of ElectroWeak Symmetry Breaking (EWSB), in fact, it is absolutely crucial to know if this new  particle is exactly the Higgs boson predicted by the SM. This goal can be achieved carefully studying the couplings of the Higgs with gauge bosons and fermions; 
in the theoretical framework of the SM the values of these couplings are fixed, and therefore any possible discrepancy can be interpreted as an undeniable signal of new physics.\\
Remarkably, the value of the Higgs mass $m_h$ measured by the two experiments ($m_h=125.2\pm 0.65$ GeV by CMS and $m_h=126.2\pm 0.67$ GeV by ATLAS) probably represents the best-suited value for this kind of analysis, allowing to investigate all the relevant final state of its decay: $b\overline{b}$, $\tau^+\tau^-$, $WW^*$, $ZZ^*$ and $\gamma\gamma$.\\
In particular a first glimpse \cite{Giardino:2012dp} to the experimental data shows an excess in the diphoton channel, confirming an indication already contained in the 2011 data \cite{ATLAS:2012ae,Giardino:2012ww}: the branching ratio of the Higgs boson into two photons seems to be around two times higher with respect to its SM value.\\
Even if it is too premature to jump to hasty conclusions \cite{Baglio:2012et}, it is worth to analyze this experimental evidence
from a theoretical point of view; the decay  $h\to\gamma\gamma$ in fact is a loop-induced process, and as a consequence it could be particularly sensitive to the presence of any new charged massive particles in addition to those belonging to the SM spectrum (see for instance \cite{calderone}).\\
As well known in the SM the decay $h\to\gamma\gamma$ is dominated by two contributions, 
the first one being induced by the (spin-$1$) gauge boson $W^{\pm}$, the other by the (spin-$1/2$) top quark; 
these two contributions are completely determined by the values of the mass, the electric charge and the spin of the particle running in the loop: the first two set the strength of the couplings, the last the underlying Lorentz structure of the interactions. This means that it is possible to define general expressions describing the diphoton decay of the 
Higgs through a spin-$1/2$ or a spin-$1$ loop. Considering as a paradigm the role of the sfermions in the Minimal Supersymmetric Standard Model (MSSM), also the loop described by a scalar (spin-0) particle has been studied.
Using these formulae it is relatively simple, therefore, to compute the effects of new particles with spin-$0$, spin-$1/2$ and spin-$1$ in the diphoton decay width of the Higgs, enhancing or decreasing each contribution with a suitable choice of their masses and/or couplings.\\
Nevertheless it can be somehow instructive to see this problem from a complementary perspective, asking oneself if the introduction of a higher-spin structure can lead to a completely different and original result.
Pursuing this idea, the aim of this paper is to analyze the effect of a spin-2 particle in the diphoton decay width of the Higgs.\\
As a theoretical framework in which this kind of contribution could play a role, 
we have in mind the possibility that the Higgs boson is not an elementary particle, rather a composite state emerging from a strongly-coupled fundamental theory \cite{Dugan:1984hq,Agashe:2004rs}.
If it were the case, in fact, a new strongly-coupled sector responsible for the EWSB
could produce the analogous of the $a_2$ meson in QCD \cite{Nakamura:2010zzi}, a charged massive resonance with $J^{PC}=2^{++}$.\\
This paper is organized as follows. In Section \ref{sec:LowEnergyTheorem} we analyze the contribution of a spin-$2$ particle to the diphoton decay width of the Higgs using a low-energy effective description; this simple description allows us to obtain in a straightforward way a first quantitative indication about the importance of such a state. In Section \ref{sec:CalcoloCompleto} we present a complete one-loop calculation, generalizing the result of Section \ref{sec:LowEnergyTheorem} to arbitrary values of the mass of the spin-2 particle. Section \ref{sec:Conclusions} is left to conclusions. In Appendix \ref{app:A} we review the basic proprieties of a charged massive spin-$2$ field, collecting all the relevant Feynman rules.

\section{The effective diphoton Higgs coupling and the spin-2 contribution}\label{sec:LowEnergyTheorem}

In this Section we analyze the effects of a spin-2 particle in the diphoton decay width of the Higgs
 using an effective field theory prescription. 
 In Section \ref{sec:2.1} we review the basic formalism, while in Section \ref{subsec:spin2} we give the details of the calculation.

\subsection{Higgs boson low-energy theorems}\label{sec:2.1}

The decay of the Higgs $h$ into two photons is a loop-induced process, 
since the Higgs boson is a neutral particle and the photon is massless. In the SM the most important contributions to this loop come from the $W^{\pm}$ boson and the top quark, while in the MSSM  there are additional scalar contributions due to sfermions and the charged Higgs.\\
In full generality we can express the diphoton decay width of the Higgs in terms of its coupling
with the particles running in the loop \cite{Djouadi:2005gi}
\begin{equation}\label{eq:hgg}
\Gamma(h\to \gamma\gamma)=\frac{\alpha^2 m_{h}^3}{1024 \pi^3}\left|
\frac{g_{hVV}}{m_V^2}Q_{V}^2A_{1}(\tau_V)+
\frac{2g_{hf\overline{f}}}{m_f}N_{c,f}Q_f^2A_{1/2}(\tau_f)+\frac{g_{hSS}}{m_S^2}N_{c,S}Q_{S}^2A_{0}(\tau_S)
\right|^2~,
\end{equation}
where $\alpha\equiv e^2/4\pi$, $m_h$ is the Higgs mass, and the indices $S$, $f$, $V$ refer, respectively, to spin-0, spin-1/2 and spin-1 particles with charges $Q_S$, $Q_f$, $Q_V$, masses $m_S$, $m_f$, $m_V$,  and couplings $g_{hSS}$, $g_{hf\overline{f}}$, $g_{hVV}$. $N_{c,f}$ and $N_{c,S}$ refer to the numbers of fermion and scalar colors while as usual we define $\tau_i\equiv 4m_i^2/m_h^2$. In the limit in which the mass of the particle running in the loop is much heavier than the mass of the Higgs the loop functions $A_{0,1/2,1}$ \cite{Djouadi:2005gi} approach the following saturating values
\begin{equation}\label{eq:numerilli}
A_{0}\to \frac{1}{3}~,~~~~~~A_{1/2}\to \frac{4}{3}~,~~~~~~A_{1}\to -7~.
\end{equation}
These values can be understood by virtue of the low-energy theorems for Higgs boson interactions \cite{Ellis:1975ap,Carena:2012xa}. These
theorems relate the amplitude of two processes which differ by the insertion of a zero momentum Higgs boson; in this limit, in fact, $h$ is a constant field and as a consequence its interactions can be obtained redefining all mass parameters of the theory in the following way
\begin{equation}\label{eq:minsub}
m_{i}~\to~m_i\left(1+g_i\,\frac{h}{v}\right)~,
\end{equation}
where $g_i$ are dimensionless numbers and  $v$ is the vacuum expectation value of the Higgs field, $v=246$ GeV. Eq. (\ref{eq:minsub}) can be immediately understood considering for definiteness the following Higgs interactions  
\begin{equation}\label{eq:IntEs}
\mathcal{L}_{int}=
-\left(1+g_S\,\frac{h}{v}\right)m_S^2\,S^2
-\left(1+g_f\,\frac{h}{v}\right)m_f\overline{f}f
-\left(1+g_V\,\frac{h}{v}\right)m_V^2V_{\mu}V^{\mu}
~,
\end{equation} 
and implies as a direct consequence the relation
\begin{equation}\label{eq:LowEnergyTheorem}
\lim_{p_h\to 0}\mathcal{M}(h\,X)=\frac{1}{v}\left(
g_S\,m_S\frac{\partial}{\partial m_S}+
g_f\,m_f\frac{\partial}{\partial m_f}+
g_V\,m_V\frac{\partial}{\partial m_V}
\right)\mathcal{M}(X)
~,
\end{equation}
where the zero four-momentum limit in the left-hand side means in practice that all particles must be considered heavy as compared with the Higgs mass. In Eq. (\ref{eq:LowEnergyTheorem}) $\mathcal{M}(h\,X)$ denotes the amplitude of a generic process involving the Higgs as an external state. For the SM case, the $W^{\pm}$ boson and top quark
couplings to the Higgs in Eq. (\ref{eq:IntEs}) are given by $g_{W,t}=1$.\\
From Eq. (\ref{eq:LowEnergyTheorem}) it follows that the $h\gamma\gamma$ interaction in the soft limit is related to the transition amplitude $\mathcal{M}(\gamma\to \gamma)$, which is just the photon two-point function. More precisely, the effect of heavy particle loops is to add the following piece to the effective QED Lagrangian density
\begin{equation}
\delta\mathcal{L}=
-\frac{1}{4}F_{\mu\nu}F^{\mu\nu}\sum_{i=0,1/2,1} \frac{\alpha}{4\pi} b_i \ln\frac{\Lambda^2}{m_i^2}~,
\end{equation}
where $\Lambda$ is an ultraviolet cutoff and $b_i$ are the beta function coefficients
\begin{equation}\label{eq:betastandard}
b_0=\frac{1}{3}N_{c,S}Q_S^2~,~~~~~~b_{1/2}=\frac{4}{3}N_{c,f}Q_f^2~,~~~~~~b_1=-7~.
\end{equation}
Using Eq. (\ref{eq:LowEnergyTheorem}) we find the following  Lagrangian density for the effective $h\gamma\gamma$ interaction
\begin{equation}\label{eq:effectiveLH}
\mathcal{L}_{h\gamma\gamma}=+\frac{\alpha}{8\pi}F_{\mu\nu}F^{\mu\nu}\frac{h}{v}\left(
g_S\,b_0+
g_f\,b_{1/2}+g_V\,b_1
\right)~;
\end{equation}
introducing the Higgs couplings $2g_V/v\equiv g_{hVV}/m_V^2$, $g_f/v\equiv g_{hf\overline{f}}/m_f$, $2g_S/v\equiv g_{hSS}/m_S^2$ this effective Lagrangian gives exactly the same result obtained  in Eqs. (\ref{eq:hgg},~\ref{eq:numerilli}) for the Higgs decay width $\Gamma(h\to \gamma\gamma)$.

\subsection{The role of a charged spin-2 resonance}\label{subsec:spin2}

\begin{figure}[!htb!]
    \centering
   \includegraphics[scale=0.95]{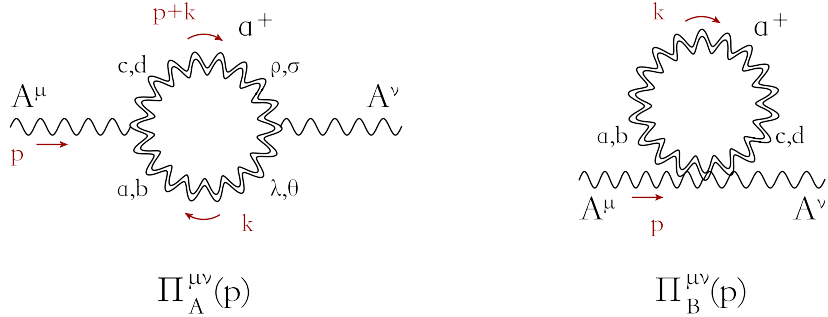}
   \caption{\textit{Relevant Feynman diagrams involved in the computation of the QED beta function coefficient $b_2$ due to the presence of a charged spin-2 resonance $a^{\pm}$.}}\label{fig:QEDbeta}
 \end{figure}

Following the perspective outlined in Section \ref{sec:2.1} we analyze here the contribution
 of a charged spin-2 resonance to the diphoton decay width $\Gamma(h\to \gamma\gamma)$. We consider a theoretical framework in which a charged spin-2 field arises as a component of a tensor isotriplet belonging to the resonances of a new strongly-interacting sector responsible for the EWSB \cite{Dugan:1984hq,Agashe:2004rs,Gripaios:2009pe}.\\
  A spin-2 isotriplet is described by a symmetric, transverse and traceless tensor matrix $\textbf{a}_{\mu\nu}$ \cite{Beyer:2006hx}
\begin{equation}\label{eq:Pro1}
\textbf{a}_{\mu\nu}=\textbf{a}_{\nu\mu}~,\hspace{1 cm}\partial^{\mu}\textbf{a}_{\mu\nu}=0~,\hspace{1 cm}\textbf{a}^{\mu}_{~\mu}=0~,
\end{equation}
\begin{equation}\label{eq:Pro2}
\textbf{a}_{\mu\nu}\equiv \sqrt{2}\left(a_{\mu\nu}^{+}\tau^{+}+a_{\mu\nu}^{-}\tau^{-}\right)
+a_{\mu\nu}^{0}\tau^{3}=
\left(
\begin{array}{cc}
  a_{\mu\nu}^{0} &  \sqrt{2}a_{\mu\nu}^{+}    \\
\sqrt{2} a_{\mu\nu}^{-}  &  -a_{\mu\nu}^{0}    
\end{array}
\right)~,
\end{equation}
where $\tau^{1,2,3}$ are the Pauli matrices and $\tau^{\pm}\equiv (\tau^{1}\pm i \tau^2)/2$.\\
 On a general ground the role of this resonance is described in terms of an effective field theory which has to respect the low-energy symmetries.
In its minimal realization this effective Lagrangian should contain a kinetic term and the lowest order in a derivative expansion of couplings to SM fermions, Higgs and  $W/Z$ bosons. A few comments are mandatory.\\ On the one hand - considering the couplings with SM fermions in the form ${\rm Tr}\left[\textbf{a}_{\mu\nu}\textbf{j}_f^{\mu\nu}\right]$ - we notice that dimension-4 couplings are forbidden, since the $\sigma^{\mu\nu}=i[\gamma^{\mu},\gamma^{\nu}]/2$ term is ruled out in the fermionic current $\textbf{j}_f^{\mu\nu}$ by the symmetric propriety in Eq. (\ref{eq:Pro1}). As a consequence the lowest order interaction is dimension-5 and must be derivative in the fermion fields; to date for these interactions no strong bounds exist.\footnote{Interestingly, in \cite{Grinstein:2012pn} the anomalously large top quark forward-backward asymmetry can naturally be accommodated in models with flavor-violating couplings of a new massive spin-2 state to quarks.}
Notice that this situation is radically 
different with respect to the case of an extra $W^{\prime}$ gauge boson. Direct couplings with SM quarks and leptons, in fact, are strongly constrained by both LHC and Tevatron searches, resulting in TeV-scale bounds on the $W^{\prime}$ mass \cite{Aad:2011yg}. In order to avoid these bounds the existence of a new $Z_2$ parity must be invoked.\\
On the other side couplings with longitudinal gauge bosons could be very interesting  from a phenomenological point of view, because they can lead to scattering amplitudes growing with the energy in the $W_LW_L$ channel \cite{Alboteanu:2008my}. Since we are dealing with an isotriplet, however, these couplings in the effective Lagrangian are forbidden if we require custodial symmetry \cite{Alboteanu:2008my}.\\
Summarizing our massive spin-2 isotriplet is described by the kinetic Lagrangian (containing its interactions with transverse gauge bosons encoded in the covariant derivatives) and its coupling with the Higgs field through the mass term. This is all we need for the calculation of the Higgs decay width $\Gamma(h\to \gamma\gamma)$. \\
In more details the Lagrangian describing a massive spin-2 field is the Fierz-Pauli Lagrangian, and in Appendix \ref{app:A} we review some of its fundamental proprieties collecting all the relevant Feynman rules, including interactions with the Higgs.\\
Here we are interested in the effective limit of the diphoton Higgs decay; following how discussed in Section \ref{sec:2.1}, therefore, this means that we need to compute the value of the QED beta function coefficient $b_2$. As well known this coefficient can be extracted from the QED two-point function of the photon, considering the loops involving the spin-2 charged resonance. Feynman diagrams are shown in Fig. \ref{fig:QEDbeta}. In full generality gauge invariance implies that the contribution of the spin-2 resonance to the two-point function of the photon must obay to the following transverse projection
\begin{equation}
i\Pi^{\mu\nu}(p)=-ip^2\left(g^{\mu\nu}-\frac{p^{\mu}p^{\nu}}{p^2}\right)\Pi^{(2)}_{\gamma\gamma}(p^2)~,
\end{equation}
 and the beta function coefficient $b_2$ can be easily read from the UV behavior of the form factor $\Pi^{(2)}_{\gamma\gamma}(p^2)$ in the zero momentum limit
 \begin{equation}\label{eq:Spin2beta}
\left.\Pi^{(2)}_{\gamma\gamma}(0)\right|_{\rm UV} = \frac{\alpha}{4\pi} b_2 \ln\frac{\Lambda^2}{m_T^2}~.
 \end{equation}
 Using the Feynman rules collected in Appendix \ref{app:A} we find
\begin{eqnarray}
i\Pi_{\rm A}^{\mu\nu}(p)&=&g^2s_W^2\int\frac{d^4 k}{(2\pi)^4}\frac{
B_{cd,\rho\sigma}(p+k)B_{\lambda\theta,ab}(k)}{\left[(p+k)^2-m_T^2\right](k^2-m_T^2)}\times\nonumber\\
&&V_{3}(k,a,b;-p-k,c,d;-p,\mu)V_3(p+k,\rho,\sigma;-k,\lambda,\theta;p,\nu)~,\label{eq:PiA}\\
&&\nonumber\\
i\Pi_{\rm B}^{\mu\nu}(p)&=&g^2s_W^2\int\frac{d^4 k}{(2\pi)^4}\frac{B_{ab,cd}(k)}{(k^2-m_T^2)}
V_4(a,b;c,d;\mu,\nu)~,\label{eq:PiB}
\end{eqnarray}
where $m_T$ is the mass of the spin-2 particle\footnote{Following the notation introduced in Eq. (\ref{eq:hgg}) we use the subscript {\tiny{$T$}} referred to the \textit{Tensorial} structure of the spin-2 particle.} and $e=gs_W$, with $s_W\equiv \sin\theta_W$, being $\theta_W$ the Weinberg angle;
in Eqs. (\ref{eq:PiA},~\ref{eq:PiB}) $B_{\mu\nu,\rho\sigma}$ describes the propagation of the spin-2 field [Eq. (\ref{eq:prop})] while $V_3$ and $V_4$ encompass the Lorentz structure of the interaction 
vertices [Eqs. (\ref{eq:V3},~\ref{eq:V4})]. Using standard reduction techniques \cite{Passarino:1978jh} we find the following expression for the transverse form factor $\Pi^{(2)}_{\gamma\gamma}(p^2)$
\begin{equation}\label{eq:TransverseFormFactor}
\Pi^{(2)}_{\gamma\gamma}(p^2)=\frac{\alpha}{2592\pi m_T^6}\frac{1}{p^2}\left[
3\mathcal{P}_{1}(p^2) B_{0}(p^2,m_T^2,m_T^2)+6\mathcal{P}_2(p^2)
A_{0} (m_T^2) + \mathcal{P}_3(p^2) \right]~,
\end{equation}
where in $D=4-2\epsilon$ dimensions
\begin{eqnarray}
A_0(m_T^2)&=& \frac{(2\pi\mu)^{2\epsilon}}{i\pi^2}\int d^Dk~ \frac{1}{(k^2-m_T^2)}~,\label{eq:A0}\\
B_0(p^2,m_T^2,m_T^2)&=&\frac{(2\pi\mu)^{2\epsilon}}{i\pi^2}\int d^Dk~ \frac{1}{\left[(p+k)^2-m_T^2\right](k^2-m_T^2)}~,\label{eq:B0}
\end{eqnarray}
and where we define the following polynomials
\begin{eqnarray}
\mathcal{P}_1(p^2)&\equiv& -1440\,m_T^8+480\,m_T^6p^2+46\,m_T^4p^4-43\,m_T^2p^6+6\,p^8~,\label{eq:poly1}\\
\mathcal{P}_2(p^2)&\equiv& +720\,m_T^6+75\,m_T^4p^2 +37\,m_T^2 p^4 - 6\,p^6~,\label{eq:poly2}\\
\mathcal{P}_3(p^2)&\equiv& -4320\,m_T^8 + 3105\,m_T^6 p^2- 804\,m_T^4 p^4 +106\,m_T^2 p^6 -6\,p^8~.\label{eq:poly3}
\end{eqnarray}
Extracting the UV part from Eqs. (\ref{eq:A0},~\ref{eq:B0}) and using Eq. (\ref{eq:Spin2beta}) we find
\begin{equation}\label{eq:b2}
b_2 = +\frac{35}{12}~.
\end{equation}
Comparing with Eq. (\ref{eq:betastandard}) we see that the QED beta function coefficient $b_2$ due to a spin-2 resonance has the opposite sign with respect to the spin-1 contribution of the $W^{\pm}$ gauge boson.\\ Using Eqs. (\ref{eq:hgg},~\ref{eq:numerilli},~\ref{eq:effectiveLH}) we can include the presence of a spin-2 particle with mass $m_T$, coupling $g_{hTT}$ and generic charge and color $Q_T$, $N_{c,T}$ in the asymptotic formula for the diphoton Higgs decay width
\begin{eqnarray}\label{eq:hgg2}
&&\Gamma(h\to \gamma\gamma)_{m_i\gg m_h}=\nonumber\\&&\frac{\alpha^2 m_{h}^3}{1024 \pi^3}\left|
\frac{g_{hTT}}{m_T^2}N_{c,T}Q_{T}^2\left(\frac{35}{12}\right)
+
\frac{g_{hVV}}{m_V^2}Q_{V}^2(-7)+
\frac{2g_{hf\overline{f}}}{m_f}N_{c,f}Q_f^2\left(\frac{4}{3}\right)
+\frac{g_{hSS}}{m_S^2}N_{c,S}Q_{S}^2\left(\frac{1}{3}\right)
\right|^2~.\nonumber\\&&
\end{eqnarray}
As a consequence of Eq. (\ref{eq:b2}) in order to obtain a constructive interference between the spin-2 and the spin-1 contributions we need a negative value for the coupling $g_{hTT}$.\\
Eq. (\ref{eq:hgg2}) contains a first quantitative information about the effect of a spin-2 particle as compared to lower spin ones.  Nevertheless the only way to achieve a complete description valid for arbitrary values of the spin-2 mass is to perform a full computation.

\section{Towards a complete calculation}\label{sec:CalcoloCompleto}

\begin{figure}[!htb!]
    \centering
   \includegraphics[scale=0.95]{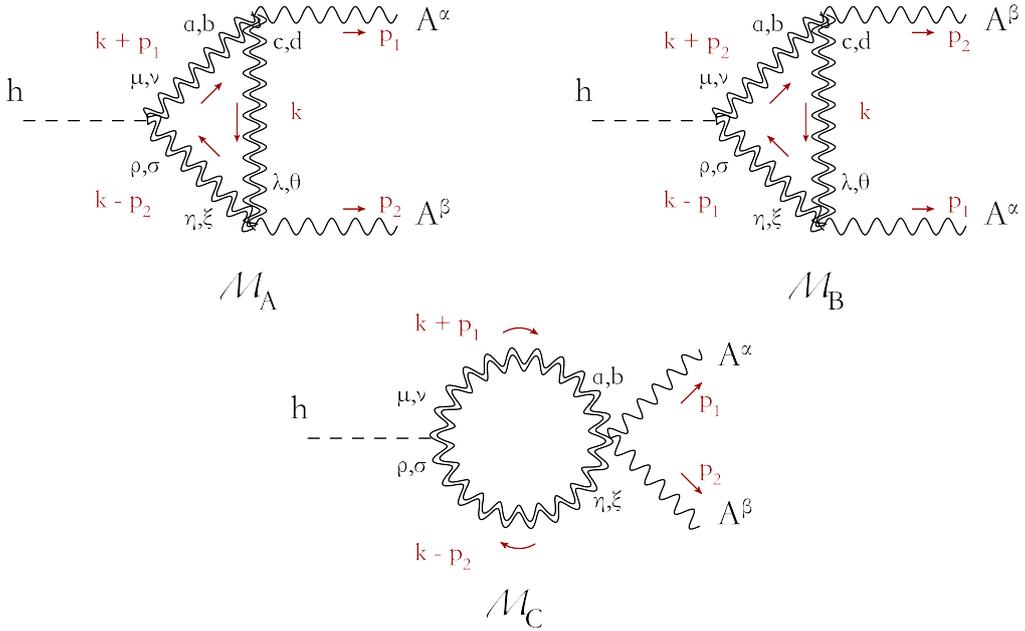}
   \caption{\textit{Feynman diagrams describing the decay of the Higgs into two photons mediated by a charged spin-2 field.}}\label{fig:Higgs}
 \end{figure}

In this Section we address the complete calculation of the diphoton Higgs decay width $\Gamma(h\to\gamma\gamma)$ mediated by the spin-2 field. Feynman diagrams are shown in Fig. \ref{fig:Higgs}. Using the Feynman rules in Fig. \ref{fig:FR} we find
\begin{eqnarray}
i\mathcal{M}_A&=& v\tilde{g}_{hTT}g^2s_W^2\int\frac{d^4k}{(2\pi)^4}\epsilon_{\alpha}^{*}(p_1)\epsilon_{\beta}^{*}(p_2)(g^{\mu\rho}g^{\nu\sigma}-g^{\mu\nu}g^{\rho\sigma})\times\nonumber\\
&&\frac{B_{\mu\nu,ab}(k+p_{1})
B_{cd,\lambda\theta}(k)
B_{\eta\xi,\rho\sigma}(k-p_{2})}
{\left[(k+p_{1})^2-m_T^2\right]\left[(k-p_{2})^2-m_T^2\right](k^2-m_T^2)}\times\nonumber\\
&&V_{3}(k+p_{1},a,b;-k,c,d;p_1,\alpha)
V_3(k,\lambda,\theta;-k+p_{2},\eta,\xi;p_2,\beta)~,\label{eq:MA}\\
&&\nonumber\\
i\mathcal{M}_C&=& v\tilde{g}_{hTT}g^2s_W^2\int\frac{d^4k}{(2\pi)^4}\epsilon_{\alpha}^{*}(p_1)\epsilon_{\beta}^{*}(p_2)(g^{\mu\rho}g^{\nu\sigma}-g^{\mu\nu}g^{\rho\sigma})\times\nonumber\\
&&\frac{B_{\mu\nu,ab}(k+p_{1})B_{\eta\xi,\rho\sigma}(k-p_{2})}{\left[(k+p_{1})^2-m_T^2\right]
\left[(k-p_{2})^2-m_T^2\right]}V_4(a,b;\eta,\xi,\alpha,\beta)~,\label{eq:MB}\\&&\nonumber
\end{eqnarray}
with $p_1^2=p_2^2=0$ and $2p_1\cdot p_2 = m_h^2$;
the expression for the amplitude $\mathcal{M}_B$ in Fig. \ref{fig:Higgs} can be straightforwardly obtained  from Eq. (\ref{eq:MA}) with the substitution $(p_1,\alpha) \leftrightarrow (p_2,\beta)$.\\
Even if this decay amplitude involves only three-point and two-point functions its reduction in terms of scalar integrals is anything but simple, due to the complicated  structure of propagators and vertices from which cumbersome tensor integrals arise.\footnote{Similar integrals can be found, e.g., in \cite{Karg:2009xk}.}\\
Nevertheless it is possible to supervise the correctness of our result using some consistency checks;
in particular the Ariadne's thread of this calculation lies in the following points.
 \begin{enumerate}
 \item 
\underline{QED Ward Identity}; replacing $\epsilon^{*}_{\alpha}(p_1)$ with $p_{1,\alpha}$ or 
$\epsilon^{*}_{\beta}(p_2)$ with $p_{2,\beta}$ the sum of the three amplitudes must give zero. This means that gauge invariance in the form of Ward Identity forces the total amplitude to be proportional to the factor
\begin{equation}\label{eq:WI}
\mathcal{M}_{A}+\mathcal{M}_B+\mathcal{M}_C \propto (p_1\cdot p_2\,g^{\alpha\beta}-p_{2}^{\alpha}p_{1}^{\beta})
\epsilon^{*}_{\alpha}(p_1)\epsilon^{*}_{\beta}(p_2)~.
\end{equation}
 
 \item
 \underline{UV divergences}; 
 using dimensional regularization and applying the usual Passarino-Veltman decomposition \cite{Denner:2005nn}  it is possible to reduce Eqs. (\ref{eq:MA},~\ref{eq:MB}) to a combination of the scalar integrals $A_0(m_T^2)$ [Eq. (\ref{eq:A0})], $B_0(p^2,m_T^2,m_T^2)$ [Eq. (\ref{eq:B0})] and $C_0(a)$, where
 \begin{equation}
 C_0(a)=\frac{(2\pi\mu)^{2\epsilon}}{i\pi^2}\int d^Dk~ \frac{1}{\left[(k+p_1+p_2)^2-m_T^2\right]
 \left[(k+p_1)^2-m_T^2\right](k^2-m_T^2)}~,
 \end{equation}
  with argument $a\equiv p_1^2,p^2_2,(p_1+p_2)^2,m_T^2,m_T^2,m_T^2$;
  bearing in mind the relation \cite{Denner:1991kt} $B_{0}(0,m_T^2,m_T^2)=A_0(m_T^2)/m_T^2-1$,  we expect in full generality to obtain the following structure
 \begin{eqnarray}\label{eq:generalUV}
&& \mathcal{M}_{A}+\mathcal{M}_B+\mathcal{M}_C \propto
 (p_1\cdot p_2\,g^{\alpha\beta}-p_{2}^{\alpha}p_{1}^{\beta})
\epsilon^{*}_{\alpha}(p_1)\epsilon^{*}_{\beta}(p_2)\times\nonumber\\
&& \left[a_1 C_0(0,0,m_h^2,m_T^2,m_T^2,m_T^2)
+a_2A_0(m_T^2)+a_3B_0(m_h^2,m_T^2,m_T^2)
+a_4
 \right]~,
 \end{eqnarray}
  with appropriate coefficients $a_{1,2,3,4}$. In the SM the  loop amplitude $h\to \gamma\gamma$ is UV-finite, 
 as the Higgs in the SM has no tree-level coupling with photons  and the theory is renormalizable. Notice that a UV-finite amplitude would require in Eq. (\ref{eq:generalUV}) the special combination $a_2=-a_3/m_T^2$, and in particular in the context of the SM both the loop involving the $W^{\pm}$ gauge boson and the top quark have $a_2=a_3=0$.\\
 This simple picture is no longer true in presence of a  charged massive spin-2 resonance whose Lagrangian, as reviewed in Appendix \ref{app:A}, is intrinsically non-renormalizable. As a consequence UV divergences will survive in the final result reflecting a unavoidable sensitivity to the UV completion of the theory. As we shall see these UV-logarithmically enhanced terms can be estimated [see Eqs. (\ref{eq:SIA0},~\ref{eq:SIB0})] switching from dimensional regularization to cutoff regularization
 \begin{equation}
 \frac{1}{\epsilon}-\gamma_E+\ln 4\pi+\ln\frac{\mu^2}{m_T^2}~~\to~~\ln\frac{\Lambda^2}{m_T^2}~.
 \end{equation}
Considering  the existence of a new strongly-coupled fundamental sector in the context of composite Higgs theories, a natural cutoff for these UV divergences is represented by the compositeness scale $\Lambda\sim$ few TeV.  
  
 \item
\underline{High-mass limit};  the limit in which the mass of the spin-2 particle inside the loop is much larger as compared to the mass of the Higgs must reproduce the result obtained in Section \ref{subsec:spin2}.

 \end{enumerate} 
 
 Pursuing the prescription outlined in these points we find the following result for the spin-2 contribution to the diphoton decay width of the Higgs
 \begin{eqnarray}
 \Gamma(h\to \gamma\gamma)_{\rm s=2}&=&\frac{\alpha^2 m_h^3}{1024 \pi^3}
 \left|
 \frac{\tilde{g}_{hTT}v}{810\,m_h^4\,\tau^5}
 \left[45m_h^4\,\mathcal{P}_{C}(\tau)\,C_0(0,0,m_h^2,m_T^2,m_T^2,m_T^2)\frac{}{}+\right.\right.\nonumber\\&&
  \left.\left. \frac{}{}2m_h^2\,\mathcal{P}_B(\tau)\,B_0(m_h^2,m_T^2,m_T^2)+\mathcal{P}_A(\tau)\,A_0(m_T^2)+2m_h^2\,
  \mathcal{P}_0(\tau)
  \right]\right|^2~,\label{eq:Gamma2}
 \end{eqnarray}
 where we define the polynomials
 \begin{eqnarray}
 \mathcal{P}_{C}(\tau)&=&\tau^2 \left\{\tau \left[3\,\tau  \left(60\,\tau ^2-95\,\tau +136\right)-268\right]+
 64\right\}~,\\
  \mathcal{P}_{B}(\tau)&=& 15\tau  \left\{\tau  \left[3\,\tau  \left(405\,\tau -703\right)+1700\right]-664\right\}+1920~,\\
  \mathcal{P}_{A}(\tau)&=& -120 \left\{\tau  \left[3\,\tau  \left(405\,\tau +2\right)+140\right]+64\right\}~,\\
  \mathcal{P}_{0}(\tau)&=& 9 \tau^2  \left\{5\,\tau  \left[15\,\tau  \left(12\,\tau +5\right)+371\right]-1076\right\}+3640\,\tau+800~,
 \end{eqnarray}
 with $\tau\equiv 4m_T^2/m_h^2$. In Appendix \ref{app:B} we collect the analytical expressions of the scalar functions $A_0$, $B_0$ and $C_0$ appearing in Eq. (\ref{eq:Gamma2}); using these formulae it is easy to check that the combination $[2m_h^2\,\mathcal{P}_B(\tau)\,B_0(m_h^2,m_T^2,m_T^2)+\mathcal{P}_A(\tau)\,A_0(m_T^2)]/\tau^5$ in Eq. (\ref{eq:Gamma2}) is UV-finite at order $\mathcal{O}(1/\tau)$, and that at the same order the decay width reproduces the result obtained in Eq. (\ref{eq:hgg2}) considering the limit $m_T\gg m_h$ and defining $g_{hTT}\equiv \tilde{g}_{hTT}v$.\\ 
On the contrary UV divergences arise in the amplitude at order $\mathcal{O}(1/\tau^2)$, reflecting the bad behavior of the spin-2 propagator for large momenta which goes like $\sim p^2/m_T^4$ thus invalidating the usual power counting arguments \cite{WeinbergBible} to deduce the renormalizability properties of a theory.\\
Due to the non-renormalizable nature of a charged massive spin-2 particle in the following we use instead of Eq. (\ref{eq:Gamma2}) its leading logarithmical (LL) approximation
\begin{equation}\label{eq:LLDecayWidth}
\Gamma(h\to \gamma\gamma)_{\rm s=2}^{\rm LL}=\frac{\alpha^2m_h^3}{1024\pi^3}
\left|\frac{\tilde{g}_{hTT}v}{m_h^2}
\left[\frac{35}{3\tau}-\frac{\mathcal{P}(\tau)}{27\tau^5}\ln\frac{\Lambda^2}{m_T^2}\right]
\right|^2~,
\end{equation}
with 
\begin{equation}\label{eq:pol}
\mathcal{P}(\tau)\equiv \tau\left[
15\,\tau\left(
141\,\tau-104
\right)+728
\right]-128~.
\end{equation} 
The reason why we concentrate on this approximation is that
its coefficients are constrained by renormalization group equations to be a function of the lowest order Lagrangian parameters only \cite{Weinberg:1978kz}, that is, they don't depend upon the higher order Lagrangian coupling constants.\footnote{The diphoton Higgs decay width in Eq. (\ref{eq:Gamma2}) has been obtained using at one loop the lowest order Lagrangian for a spin-2 interacting particle, as discussed in Appendix \ref{app:A}. At the same perturbative order and abandoning the LL approximation one should consider also the tree level diagrams describing the process $h\to \gamma\gamma$ coming from the first order Lagrangian. In particular at dim-6 we have three different contributions
\begin{equation}
\mathcal{L}_{HV}^{6}=-\frac{c_Wg^2}{2\Lambda^2}|H|^2W_{\mu\nu}^aW^{a,\mu\nu}
-\frac{c_Bg^{\prime 2}}{2\Lambda^2}|H|^2B_{\mu\nu}B^{\mu\nu}-
\frac{c_{WB}gg^{\prime}}{2\Lambda^2}\left(H^{\dag}\tau^{a}H\right)B_{\mu\nu}W^{a,\mu\nu}~,
\end{equation} 
which, after EWSB, lead to
\begin{equation}
\mathcal{L}_{HV}^{6}=-\frac{c_{\gamma}e^2}{2\Lambda^2}\,(hv)\,F_{\mu\nu}F^{\mu\nu}+\cdots ~,
\end{equation}
with $c_{\gamma}\equiv c_W+c_B-c_{WB}$.
}
Using Eq. (\ref{eq:LLDecayWidth}) it is possible to study the effects of a spin-2 particle in the diphoton decay width of the Higgs in the LL approximation. We  address this analysis in the following section.

\section{Discussion and Conclusions}\label{sec:Conclusions}

Defining the following ratio
\begin{equation}\label{eq:ratioR}
R_{XX}\equiv \frac{\sigma(pp\to h)}{\sigma_{\rm SM}(pp\to h)}\frac{\Gamma(h\to XX)}{\Gamma_{\rm SM}(h\to XX)}~,
\end{equation}
between the observed Higgs signal in the decay channel $h\to XX$ and its SM prediction we find  \cite{Giannotti,:2012gk,Incandela,:2012gu,Baglio:2012et}
\begin{equation}\label{eq:Table}
\begin{array}{ | c | c | c |}
\hline
  & R_{\gamma\gamma}  & R_{ZZ}   \\
  \hline
 {\rm ATLAS} &  1.90\pm 0.5 & 1.3\pm0.6  \\
  {\rm CMS} &   1.56\pm 0.43  & 0.7\pm 0.5 \\
  {\rm ATLAS\,\oplus\,CMS} & 1.71\pm 0.33  &  0.95\pm 0.4  \\
  \hline
\end{array}~.
\end{equation}
From the table in Eq. (\ref{eq:Table}) we see that, while the signal strength in the channel $h\to ZZ$ seems to be in good agreement with the SM value, the biggest discrepancy is related to the diphoton decay $h\to \gamma\gamma$. In this Section we try to interpret this evidence analyzing the effects of a spin-2 particle using the results obtained in Section \ref{sec:CalcoloCompleto}. For simplicity and clearness we study the possibility to enhance the rate $R_{\gamma\gamma}$ increasing the partial diphoton decay width of the Higgs but without changing its production cross section or its total width with respect to their SM values; this means that we consider a colorless spin-2 particle, while Eq. (\ref{eq:ratioR}) simplifies to
\begin{equation}\label{eq:Ratio2}
R_{\gamma\gamma}=\frac{\Gamma(h\to \gamma\gamma)}{\Gamma_{\rm SM}(h\to \gamma\gamma)}~.
\end{equation}
 \begin{figure}[!htb!]
    \centering
   \includegraphics[scale=0.7]{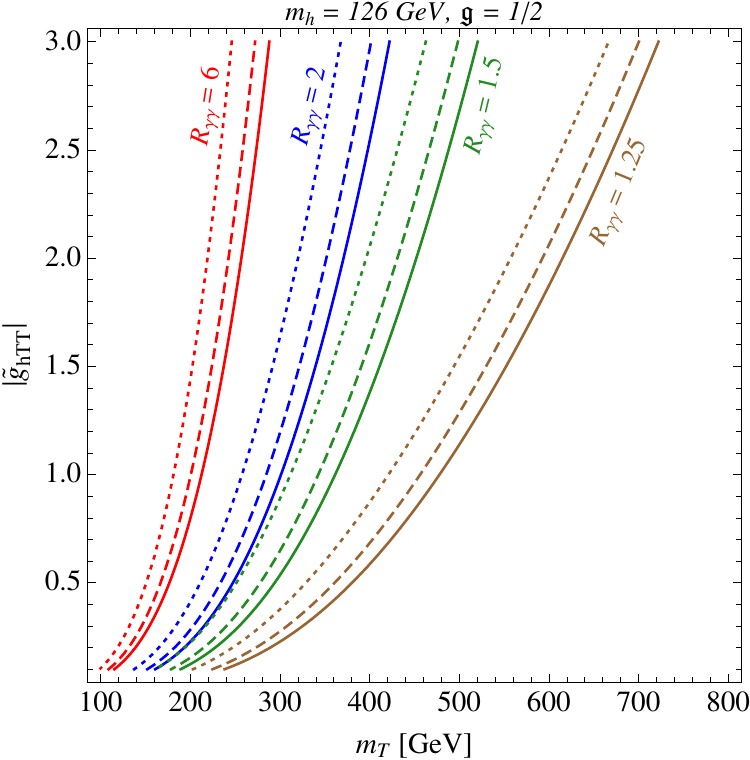}
   \caption{\textit{Contours of constant diphoton Higgs decay width normalized to the SM value [Eq. (\ref{eq:Ratio2})] in presence of a spin-2 particle. Solid, dashed and dotted lines correspond to $\Lambda=7$ TeV, $\Lambda=3$ TeV and $\Lambda=1$ TeV. We use the value} $\texttt{g}=1/2$ \textit{for the gyromagnetic ratio of the spin-2 particle (see Appendix \ref{app:A} for details and Appendix \ref{app:C} for the case} $\texttt{g}=2$\textit{).}}\label{fig:PlotR}
 \end{figure} 
In Fig. \ref{fig:PlotR} we present the value of this ratio in presence of a spin-2 
particle in the plane $(m_T, \tilde{g}_{hTT})$. We set $m_h=126$ GeV, and we show three possible values ($\Lambda=7$ TeV, $\Lambda=3$ TeV  and $\Lambda=1$ TeV) for the UV cutoff.  In particular we see that an enhancement in the rate $R_{\gamma\gamma}$ is possible  even for a small value of the coupling $|\tilde{g}_{hTT}|\sim 0.5$ considering $m_T\sim$ few hundred GeV.\\

In this work, we have analyzed the possibility to enhance the diphoton decay width of the Higgs postulating the existence of a charged massive spin-2 particle; the presence of such a resonance can be justified, in analogy with QCD, if we assume the existence at the TeV scale of a new strongly-coupled fundamental sector responsible for the EWSB. This is a completely new calculation, since this high-spin structure has never been studied in the literature in this particular  context.\\
The main phenomenological results of this paper are summarized in Eqs. (\ref{eq:hgg2},~\ref{eq:Gamma2},~\ref{eq:LLDecayWidth}) and in Fig. \ref{fig:PlotR}. In Eq.  (\ref{eq:hgg2}) we have calculated  the QED beta function coefficient  related to the presence of a charged massive spin-2 particle, which enters in the diphoton decay width considering the limit $m_T\gg m_h$; in Eq. (\ref{eq:Gamma2})  we have presented the analytical expression for the decay width $\Gamma(h\to \gamma\gamma)$ due to the loop induced by the spin-2 particle together with its logarithmical approximation in Eq. (\ref{eq:LLDecayWidth}); finally in Fig. \ref{fig:PlotR} we have shown the enhancement in the Higgs diphoton rate as compared with its SM value.\\
From a theoretical point of view we have found that the non-renormalizable nature of the Lagrangian describing a charged massive spin-2 particle leads to the presence of UV divergences in the expression of the decay width $\Gamma(h\to \gamma\gamma)$. On the one hand this is an alarm bell to keep in mind: even behind this observable, UV-finite in the context of the SM, a strong sensitivity to the UV completion of the theory could be hidden.
Due to this same UV sensitivity, on the other hand, the phenomenological usefulness of this spin-2 contribution in the diphoton  decay width of the Higgs seems to be rather limited.

\subsection*{Aknowledgement}

The author is deeply indebted to Denis Comelli for reading the manuscript, for his constant encouragements
and precious advices and to Luigi Delle Rose, Eugenio Del Nobile and Massimo Porrati for useful discussions.\\
 This work is supported by the Emergence-UPMC-2011 research program.\\

\appendix
\section{Massive spin-2 field and the Fierz-Pauli Lagrangian}\label{app:A}

\begin{figure}[!htb!]
    \centering
   \includegraphics[scale=0.9]{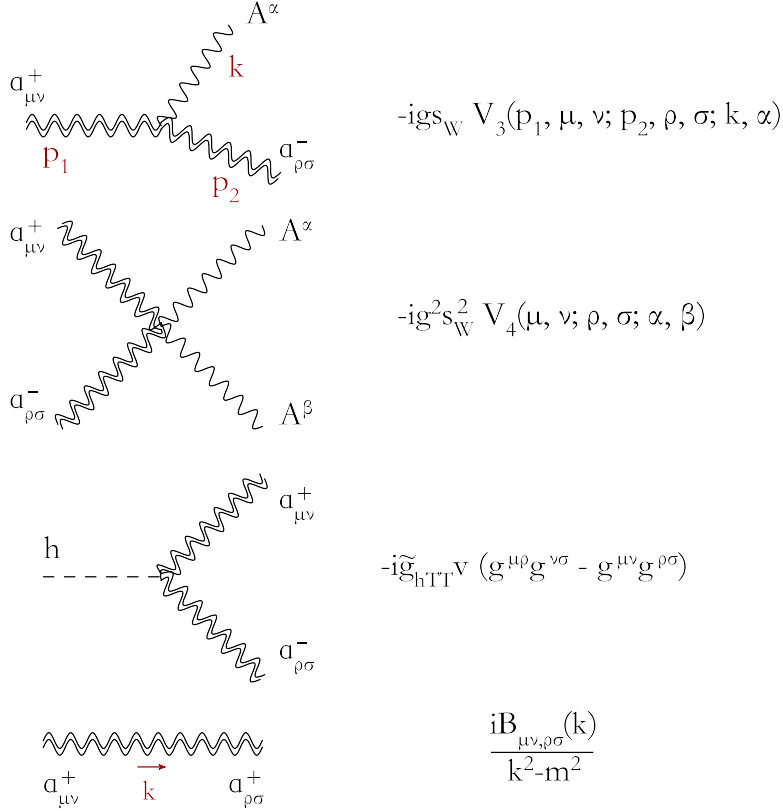}
   \caption{\textit{Feynman rules describing the propagator and the interactions of the charged spin-2 resonance $a^{\pm}$ with the photon and the Higgs; the Lorentz structures are explicitly written in Eqs. (\ref{eq:V3},~\ref{eq:V4},~\ref{eq:prop}). Four-momenta are taken to be incoming. In Sections \ref{sec:LowEnergyTheorem},~\ref{sec:CalcoloCompleto},~\ref{sec:Conclusions} we consider the value} $\texttt{g}=1/2$\textit{, while in Appendix \ref{app:C} we discuss the case} $\texttt{g}=2$\textit{.}}\label{fig:FR}
 \end{figure}

The $5$ degrees of freedom (DOF) of a massive spin-2 field are described by a rank-two symmetric, transverse and traceless tensor $a_{\mu\nu}$ 
\begin{equation}
a_{\mu\nu}=a_{\nu\mu}~,\hspace{1 cm}\partial^{\mu}a_{\mu\nu}=0~,\hspace{1 cm}a^{\mu}_{~\mu}=0~,
\end{equation}
satisfying the mass-shell condition
\begin{equation}
\left(\partial^{2}+m_0^{2}\right)a_{\mu\nu}=0~.
\end{equation}
It's possible to show that all these proprieties can be derived from the following Fierz-Pauli Lagrangian density \cite{Fierz:1939ix,Hinterbichler:2011tt}
\begin{eqnarray}\label{eq:FPL}
\mathcal{L}_{\rm FP}&=&\frac{1}{2}\left(\partial_{\mu}a_{\nu\rho}\right)\left(\partial^{\mu}a^{\nu\rho}\right)
-\left(\partial_{\mu}a_{\nu\rho}\right)\left(\partial^{\nu}a^{\mu\rho}\right)-\frac{1}{2}\left(\partial_{\mu}a\right)\left(\partial^{\mu}a\right)+\left(\partial^{\mu}a_{\mu\nu}\right)\left(\partial^{\nu}a\right)\nonumber\\&-&
\frac{m_0^{2}}{2}\left(a^{\mu\nu}a_{\mu\nu}-a^2\right)~,
\end{eqnarray}
where $a\equiv a^{\mu}_{~\mu}$.\\
In this article we are dealing with a spin-2 resonance transforming as a triplet under $SU(2)_L$; this isospin structure can be straightforwardly introduced considering the matrix
\begin{equation}
\textbf{a}_{\mu\nu}\equiv \sqrt{2}\left(a_{\mu\nu}^{+}\tau^{+}+a_{\mu\nu}^{-}\tau^{-}\right)
+a_{\mu\nu}^{0}\tau^{3}=
\left(
\begin{array}{cc}
  a_{\mu\nu}^{0} &  \sqrt{2}a_{\mu\nu}^{+}    \\
\sqrt{2} a_{\mu\nu}^{-}  &  -a_{\mu\nu}^{0}    
\end{array}
\right)~,
\end{equation}
and as a consequence - introducing a trace with respect to $SU(2)_L$ indices - the Lagrangian density in Eq. (\ref{eq:FPL}) becomes
\begin{eqnarray}\label{eq:FPmatrix}
\mathcal{L}_{\rm FP}^{SU(2)_L}&=& \frac{1}{4}~{\rm Tr}\left[\left(D_{\mu}\textbf{a}_{\nu\rho}\right)^{\dag}\left(D^{\mu}\textbf{a}^{\nu\rho}\right)\right]
-\frac{1}{2}~{\rm Tr}\left[\left(D_{\mu}\textbf{a}_{\nu\rho}\right)^{\dag}\left(D^{\nu}\textbf{a}^{\mu\rho}\right)\right]-\frac{1}{4}~{\rm Tr}\left[\left(D_{\mu}\textbf{a}\right)^{\dag}\left(D^{\mu}\textbf{a}\right)\right]\nonumber\\&+&\frac{1}{2}~{\rm Tr}\left[\left(D^{\mu}\textbf{a}_{\mu\nu}\right)^{\dag}
\left(D^{\nu}\textbf{a}\right)\right]-
\frac{m_0^{2}}{4}\left({\rm Tr}\left[\textbf{a}^{\mu\nu}\textbf{a}_{\mu\nu}\right]-{\rm Tr}\left[\textbf{a}^2\right]\right)~,
\end{eqnarray}
where for the covariant derivative we have
\begin{equation}
D_{\mu}\textbf{a}_{\nu\rho}=\partial_{\mu}\textbf{a}_{\nu\rho}+\frac{ig}{2}W_{\mu}^{a}\tau^{a}\textbf{a}_{\nu\rho}-\frac{ig^{\prime}}{2}B_{\mu}\textbf{a}_{\nu\rho}\tau^{3}~.
\end{equation}
Focusing on the electromagnetic interactions of the charged component $a^{\pm}$ we can extract from Eq. (\ref{eq:FPmatrix}) the charged counterpart of Eq. (\ref{eq:FPL})
\begin{eqnarray}\label{eq:FPLCharged}
\mathcal{L}_{\rm FP}^{\pm}&=&\left(\mathcal{D}_{\mu}a^+_{\nu\rho}\right)\left(\mathcal{D}^{\mu}a^{-,\nu\rho}\right)
-2\left(\mathcal{D}_{\mu}a^+_{\nu\rho}\right)\left(\mathcal{D}^{\nu}a^{-,\mu\rho}\right)
-\left(\mathcal{D}_{\alpha}a^+\right)\left(\mathcal{D}^{\alpha}a^-\right)
\nonumber\\&+&
\left(\mathcal{D}^{\mu}a_{\mu\nu}^+\right)\left(\mathcal{D}^{\nu}a^-\right)+
\left(\mathcal{D}^{\alpha}a^+\right)\left(\mathcal{D}^{\rho}a^-_{\rho\alpha}\right)
-m_0^{2}\left(a^{+,\mu\nu}a^-_{\mu\nu}-a^+a^-\right)~,
\end{eqnarray}
where we introduce the following notation for the QED covariant derivative
\begin{equation}\label{eq:QEDcov}
\mathcal{D}^{\alpha}a_{\mu\nu}^{\pm}\equiv \left(\partial^{\alpha}\pm i e A^{\alpha}\right)a_{\mu\nu}^{\pm}~.
\end{equation}
It's well known, however, that the minimal  coupling described by Eqs. (\ref{eq:FPLCharged},~\ref{eq:QEDcov}) is ambiguous, leading to a wrong DOF count for  a spin-2 particle \cite{Hagen:1972yp,Porrati}; without entering into details, it turns out that this problem can be fixed adding to Eq. (\ref{eq:FPLCharged}) the following interaction term \cite{Hagen:1972yp,Porrati}
\begin{equation}\label{eq:FederbushL}
\mathcal{L}_{\rm FP}^{\prime\,\pm} = -2ie\,\texttt{g}\,a_{\mu\nu}^{+}F^{\mu\rho}a_{\rho\sigma}^{-}g^{\nu\sigma}~,
\end{equation}
where $F^{\mu\rho}=\partial^{\mu}A^{\rho}-\partial^{\rho}A^{\mu}$ and $\texttt{g}$ is the gyromagnetic ratio; in particular it is possible to show that the choice $\texttt{g}=1/2$ is the unique one able to restore the correct number of propagating DOF. \\
Notice that even at cost of an incorrect DOF count, also the value $\texttt{g}=2$ has been proposed using the requirement of tree unitarity \cite{WeinbergBook,Ferrara:1992yc} and the observation that massive higher spin string states couple to electromagnetic background with this particular value \cite{Ferrara:1992yc}. Bearing in mind the existence of this alternative choice, in this paper we focus on the $\texttt{g}=1/2$ model, since it preserves the correct number of propagating DOF offering for our phenomenological purposes  a more reliable effective  description \cite{Deser:2001dt}. Nevertheless in Appendix \ref{app:C} we generalize our results to the $\texttt{g}=2$ model.\\ 
The Lagrangian $\mathcal{L}^{\pm}_{\rm FP}+\mathcal{L}_{\rm FP}^{\prime\,\pm}$ with $\texttt{g}=1/2$ is known in literature as the Federbush Lagrangian \cite{federbush}, and contains the electromagnetic interactions mediating the decay process $h\to \gamma\gamma$.\\ 
For completeness notice that using the gauge-invariant language of Eq. (\ref{eq:FPmatrix}) we can translate Eq. (\ref{eq:FederbushL}) into\footnote{The stability of a massive spin-2 field in presence of electroweak interactions is an open theoretical problem, and goes well beyond the phenomenological purpose of this paper since we are interested in genuine electromagnetic interactions. A further investigation will be presented in a forthcoming work \cite{mio2}.}
\begin{equation}
\mathcal{L}_{\rm FP}^{\prime\,SU(2)_L}=\frac{i\texttt{g}}{2}\,
{\rm Tr}\left[\textbf{a}^{\mu\nu}\left(gW_{\mu\rho}^{a}\tau^{a}-g^{\prime}B_{\mu\rho}\right)\textbf{a}^{\rho\sigma}\right]g_{\nu\sigma}~,
\end{equation}
where $B_{\mu\nu}=\partial_{\mu}B_{\nu}-\partial_{\nu}B_{\mu}$ and $W_{\mu\nu}^a=\partial
_{\mu}W_{\nu}^a-\partial_{\nu}W_{\mu}^{a}-g\epsilon^{abc}W_{\mu}^bW_{\nu}^c$.\\
At this point before to proceed it is necessary to do an observation of primary importance. Even if the Federbush Lagrangian leads to the propagation of the correct number of DOF for a spin-2 particle, in fact, it lacks of full theoretical consistency due to its causality difficulties \cite{Kobayashi:1978mv}. These problems can be fixed introducing higher-dimensional operators, making as a consequence the Lagrangian for a  charged massive spin-2 object intrinsically non-renormalizable \cite{Porrati:2008gv}. In this paper we work at the level of dim-4 operators, thus neglecting higher-dimensional corrections; notice however that in our perspective this pathology generates the presence of UV divergences in the computation of the diphoton decay width of the Higgs (see Section \ref{sec:CalcoloCompleto}).\\ 
Let us now turn to discuss the coupling of the Higgs boson with the charged spin-2 field $a^{\pm}$.\\
This coupling is encoded, mimicking the mass term in Eq. (\ref{eq:FPmatrix}), in the following Lagrangian density
\begin{equation}
\mathcal{L}_{hTT}=-\frac{\tilde{g}_{hTT}}{4}|H|^2\left({\rm Tr}\left[\textbf{a}^{\mu\nu}\textbf{a}_{\mu\nu}\right]-{\rm Tr}\left[\textbf{a}^2\right]\right)~,
\end{equation}
where $H$ is the SM Higgs doublet whose vacuum expectation value is $\langle H\rangle = (0,v)^T/\sqrt{2}$. After EWSB we find
\begin{equation}
\mathcal{L}_{hTT}\ni -\tilde{g}_{hTT}v\,h\left(g^{\mu\rho}g^{\nu\sigma}-g^{\mu\nu}g^{\rho\sigma}\right)a_{\mu\nu}^{+}a_{\rho\sigma}^{-}~,
\end{equation}
while the mass squared of the spin-2 resonance receives a contribution proportional to  $v^2$; we define
\begin{equation}
m_T^{2}\equiv m_0^2+\tilde{g}_{hTT}v^2/4~.
\end{equation}
All in all we can summarize the relevant interactions in the Feynman rules collected in Fig. \ref{fig:FR},
where we find the following Lorentz structures for the propagator \cite{vanDam:1970vg} and for the vertices $V_{3}$ and $V_{4}$
\begin{eqnarray}
V_{3}(p_1,\mu,\nu;p_2,\rho,\sigma;k,\alpha)&=&(p_1-p_2)_{\alpha}(g_{\mu\rho}g_{\nu\sigma}-g_{\mu\nu}g_{\rho\sigma})-2g_{\nu\sigma}(p_{1,\rho}g_{\alpha\mu}-p_{2,\mu}g_{\alpha\rho})\nonumber\\
&+&g_{\rho\sigma}(p_{1,\mu}g_{\nu\alpha}-p_{2,\nu}g_{\mu\alpha})+g_{\mu\nu}
(p_{1,\sigma}g_{\rho\alpha}-p_{2,\rho}g_{\sigma\alpha})\nonumber\\&+&2\,\texttt{g}\,g_{\nu\sigma}(k_{\mu}g_{\rho\alpha}-k_{\rho}
g_{\mu\alpha})~,\label{eq:V3}\\
V_{4}(\mu,\nu;\rho,\sigma;\alpha,\beta)&=& 2g_{\alpha\beta}(g_{\mu\nu}g_{\rho\sigma}-g_{\mu\rho}g_{\nu\sigma})+2g_{\nu\sigma}(g_{\alpha\rho}g_{\mu\beta}+g_{\mu\alpha}g_{\rho\beta})\nonumber\\
&-&g_{\rho\sigma}(g_{\mu\alpha}g_{\nu\beta}+g_{\mu\beta}g_{\nu\alpha})-g_{\mu\nu}(
g_{\sigma\alpha}g_{\rho\beta}+g_{\sigma\beta}g_{\rho\alpha})~,\label{eq:V4}\\
B_{\mu\nu,\rho\sigma}(k)&=&\frac{1}{2}P_{\mu\rho}P_{\nu\sigma}+
\frac{1}{2}P_{\mu\sigma}P_{\nu\rho}-\frac{1}{3}P_{\mu\nu}P_{\rho\sigma}~,\label{eq:prop}
\end{eqnarray}
where
\begin{equation}
P_{ab}\equiv \left(g_{ab}-\frac{k_{a}k_{b}}{m^2}\right)~.
\end{equation}

\section{Loop scalar functions}\label{app:B}

In this Appendix we collect the loop scalar functions used throughout this paper.
\begin{equation}\label{eq:SIA0}
A_0(m_T^2)=m_T^2\left(\Delta + \ln\frac{\mu^2}{m_T^2}+1\right)~,
\end{equation}
\begin{equation}\label{eq:SIB0}
B_0(m_h^2,m_T^2,m_T^2)=\left\{
\begin{array}{cc}
  \Delta+2+\ln\frac{\mu^2}{m_T^2}-2\sqrt{\tau-1}\tan^{-1}\left(\frac{1}{\sqrt{\tau-1}}\right)~, & \tau \geqslant 1~,     \\
   \Delta+2+\ln\frac{\mu^2}{m_T^2}-\sqrt{1-\tau}\left[i\pi+2\coth^{-1}\left(\frac{1}{\sqrt{1-\tau}}\right)\right]
  ~, & \tau < 1~,
\end{array}
\right.
\end{equation}
\begin{equation}
C_0(0,0,m_h^2,m_T^2,m_T^2,m_T^2)=\left\{
\begin{array}{cc}
   -\frac{2}{m_h^2}\arcsin^2\tau^{-1/2}~,   & \tau \geqslant 1~,   \\
    \frac{1}{2m_h^2}\left(\ln\frac{1+\sqrt{1-\tau}}{1-\sqrt{1-\tau}}-i\pi\right)^2~,  &   \tau<1~,
\end{array}
\right.
\end{equation}
with $\Delta\equiv \epsilon^{-1}-\gamma_E+\ln 4\pi$, being $\gamma_E$ the Euler-Mascheroni constant.

\section{Gyromagnetic ratio $\texttt{g}=2$}\label{app:C}

\begin{figure}[!htb!]
    \centering
   \includegraphics[scale=0.7]{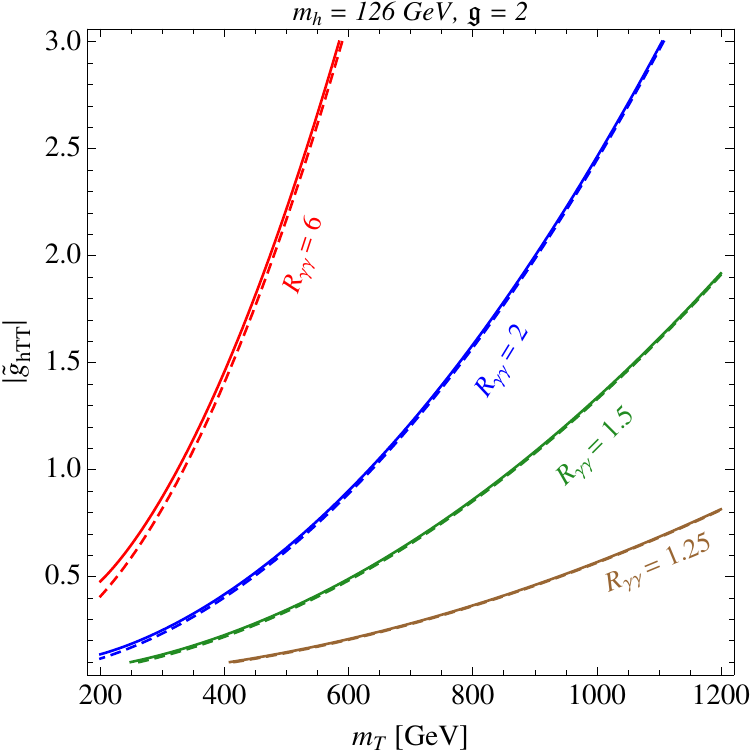}
   \caption{\textit{Contours of constant diphoton Higgs decay width normalized to the SM value [Eq. (\ref{eq:Ratio2})] in presence of a spin-2 particle with gyromagnetic ratio} $\texttt{g}=2$\textit{. Solid (dashed) lines correspond to $\Lambda=7$ TeV  
   ($\Lambda=1.5$ TeV).}}\label{fig:PlotRbis}
 \end{figure}

In this Appendix we generalize the results obtained in Sections \ref{sec:LowEnergyTheorem},~\ref{sec:CalcoloCompleto},~\ref{sec:Conclusions}. We consider a generic value $\texttt{g}$ for the gyromagnetic ratio, discussing the particular choice $\texttt{g}=2$. There are, in fact, several arguments pointing towards this  value as the one that yields a more realistic behavior for a spin-2 particle:
\begin{enumerate}
\item optical and low energy theorems, as discussed in \cite{WeinbergBook}, imply  $\texttt{g}=2$;
\item  massive higher-spin string states couple to electromagnetic backgrounds
with $\texttt{g}=2$ \cite{Ferrara:1992yc}; 
\item fundamental ${\rm spin}\,\leqslant 1$ particles observed to date
couple with $\texttt{g}=2$.
\end{enumerate}
Considering in Eq. (\ref{eq:TransverseFormFactor}) the transverse form factor $\Pi_{\gamma\gamma}^{(2)}
(p^2)$ we find 
\begin{equation}\label{eq:GenericTransverseFormFactor}
\Pi^{(2)}_{\gamma\gamma}(p^2,\texttt{g})=\frac{\alpha}{1296\pi m_T^6}\frac{1}{p^2}\left[
3\mathcal{P}_{1}(p^2,\texttt{g}) B_{0}(p^2,m_T^2,m_T^2)+6\mathcal{P}_2(p^2,\texttt{g})
A_{0} (m_T^2) + \mathcal{P}_3(p^2,\texttt{g}) \right]~,
\end{equation}
where
\begin{eqnarray}
\mathcal{P}_{1}(p^2,\texttt{g})&=&-\left(4\,m_T^2-p^2\right) \left\{60 [\texttt{g} (3\,\texttt{g}-8)+3] 
m_T^6 p^2-2 [\texttt{g} (19\,\texttt{g}-84)+42]
 m_T^4 p^4-\right.\nonumber\\&&\left.4 [\texttt{g} (\texttt{g}+4)-3] m_T^2 p^6+(1-2\,\texttt{g})^2 
 p^8+180\,m_T^8\right\}~,\\
\mathcal{P}_{2}(p^2,\texttt{g})&=& 30 [\texttt{g} (9\,\texttt{g}+16)-9] 
m_T^6 p^2-2 [\texttt{g} (23\,\texttt{g}+72)-51] 
m_T^4 p^4+\nonumber\\&&[4\,\texttt{g} (4\,\texttt{g}+1)-9] 
m_T^2 p^6-(1-2\,\texttt{g})^2 p^8+360\,m_T^8~,\\
\mathcal{P}_{3}(p^2,\texttt{g})&=& 90 [\texttt{g} (9\,\texttt{g}-8)+19] m_T^8 p^2-6 [16\,\texttt{g} (11\, \texttt{g}-8)+87] m_T^6 p^4+\nonumber\\&&[4\,\texttt{g} (110\,\texttt{g}-93)+
129] m_T^4 p^6-4 [\texttt{g} (23\,\texttt{g}-22)+6] m_T^2 p^8+\nonumber\\&&2 (1-2\,\texttt{g})
^2 p^{10}-2160\,m_T^{10}~,
\end{eqnarray}
with $\mathcal{P}_{1,2,3}(p^2,\texttt{g}=1/2)\equiv \mathcal{P}_{1,2,3}(p^2)$ in Eqs. (\ref{eq:poly1},~\ref{eq:poly2},~\ref{eq:poly3}). From Eq. (\ref{eq:GenericTransverseFormFactor}) we find using Eq. (\ref{eq:Spin2beta})
\begin{equation}
b_2(\texttt{g})=-\frac{180[(\texttt{g}-16)\texttt{g}+6]}{108}~,
\end{equation}
and as a consequence 
\begin{equation}
b_2(\texttt{g}=2)=\frac{110}{3}~.
\end{equation}
Considering in Eq. (\ref{eq:LLDecayWidth}) the diphoton decay width of the Higgs in presence of a spin-2 particle we find in LL approximation the following generalization
\begin{equation}\label{eq:GenericLLDecayWidth}
\Gamma(h\to \gamma\gamma,\texttt{g})_{\rm s=2}^{\rm LL}
=\frac{\alpha^2m_h^3}{1024\pi^3}
\left|\frac{\tilde{g}_{hTT}v}{m_h^2}
\left[-\frac{20}{3\tau}\,[(\texttt{g}-16)\texttt{g}+6]
-\frac{32\,\mathcal{P}(\tau,\texttt{g})}{27\tau^5}\ln\frac{\Lambda^2}{m_T^2}\right]
\right|^2~,
\end{equation}
with 
\begin{equation}
\mathcal{P}(\tau,\texttt{g})
\equiv \frac{45}{8} [\texttt{g} (\texttt{g}+12)-18]  \tau ^3+15 (\texttt{g}-7) (\texttt{g}-1) 
 \tau ^2+[\texttt{g}(5\,\texttt{g}+68)-58]  \tau +16 (\texttt{g}-1)^2~.
\end{equation} 
In Fig. \ref{fig:PlotRbis} we show the enhancement in the diphoton rate $R_{\gamma\gamma}$ in presence of a spin-2 particle with $\texttt{g}=2$, noticing a weaker dependence from the UV-cutoff of the theory.\\
 Considering this particular choice for the gyromagnetic ratio, we see that 
 the same contour of constant enhancement in the diphoton decay width of the Higgs can be reach for larger values  of the spin-2 mass $m_T$ and smaller values of the coupling $|\tilde{g}_{hTT}|$ with respect to the $\texttt{g}=1/2$ model in Fig. \ref{fig:PlotR}.

\newpage



\end{document}